\documentclass{article}

\usepackage[utf8]{inputenc} 
\usepackage[T1]{fontenc}
\usepackage{url}
\usepackage{ifthen}
\usepackage{cite}
\usepackage[cmex10]{amsmath} 

\usepackage{mathtools}

\usepackage{algorithmic}

\usepackage{amsfonts} % needed for \mathbb
\usepackage{extarrows} % needed for \xlongrightarrow etc.

\def\mymedskip{\vskip\medskipamount}
\def\mymedbreak{\par \ifdim\lastskip<\medskipamount
  \removelastskip \penalty-100 \mymedskip \fi}
\def\myaftermedspace{\par \ifdim\lastskip<\medskipamount
  \removelastskip \penalty55\mymedskip\fi}
\newcommand{\eop}{{\unskip\nobreak\hfil\penalty50
          \hskip2em\hbox{}\nobreak\hfil$\Box$
          \parfillskip=0pt \finalhyphendemerits=0 \par}}
\newenvironment{proof}%
{\mymedbreak{\noindent\bf Proof.\enspace}}{\eop\myaftermedspace}
\newenvironment{proofn}[1]%
{\mymedbreak{\noindent\bf Proof #1.\enspace}}{\eop\myaftermedspace}
{\mymedbreak{\noindent\bf Proof of Theorem~\ref{#1}:\enspace}}{\eop\myaftermedspace}
{\mymedbreak\noindent{\bf Remark:}%
\enspace\rm}%
{\myaftermedspace}
\newtheorem{teor}{Theorem}[section]
\newtheorem{defi}[teor]{Definition}
\newtheorem{fact}[teor]{Fact}
\newtheorem{problem}{Problem}
\newtheorem{exercise}{Exercise}
\newtheorem{examp}[teor]{Example}
\newtheorem{lem}[teor]{Lemma}
\newtheorem{cor}[teor]{Corollary}
\newtheorem{con}[teor]{Conjecture}
\newtheorem{ques}[teor]{Question}
\newtheorem{prop}[teor]{Proposition}

\newtheorem{rem}[teor]{Remark}
\newcommand{\beq}{\begin{equation}}
\newcommand{\eeq}{\end{equation}}
\newcommand{\beql}[1]{\begin{equation} \label{#1}}
\newcommand{\eeql}{\end{equation}}
\newcommand{\beqa}{\begin{IEEEeqnarray*}{rCl}}
\newcommand{\eeqa}{\end{IEEEeqnarray*}}
\newcommand{\beqal}[1]{\begin{IEEEeqnarray}{rCl} \label{#1}}
\newcommand{\eeqal}{\end{IEEEeqnarray}}
\newcommand{\beqan}{\begin{IEEEeqnarray}{rCl}}
\newcommand{\eeqan}{\end{eqnarray}}
\newcommand{\bpf}{\begin{proof}}
\newcommand{\epf}{\end{proof}}
\newcommand{\bpfn}[1]{\begin{proofn}{#1}}
\newcommand{\epfn}{\end{proofn}}
\newcommand{\ben}{\begin{enumerate}}
\newcommand{\een}{\end{enumerate}}
\newcommand{\bit}{\begin{itemize}}
\newcommand{\eit}{\end{itemize}}

\newcommand{\bab}{\begin{abstract}}
\newcommand{\eab}{\end{abstract}}
\newcommand{\bke}{\begin{keywords}}
\newcommand{\eke}{\end{keywords}}

\newcommand{\btm}[1]{\begin{teor} \label{#1}}
\newcommand{\etm}{\end{teor}}
\newcommand{\btmn}[2]{\begin{teor}[#1] \label{#2}}
\newcommand{\etmn}{\end{teor}}
\newcommand{\ble}[1]{\begin{lem} \label{#1}}
\newcommand{\ele}{\end{lem}}
\newcommand{\bLe}[1]{\begin{Lemma} \label{#1}}
\newcommand{\eLe}{\end{Lemma}}
\newcommand{\blen}[2]{\begin{lem}[#1] \label{#2}}
\newcommand{\elen}{\end{lem}}
\newcommand{\bpn}[1]{\begin{prop} \label{#1}}
\newcommand{\epn}{\end{prop}}
\newcommand{\bex}[1]{\begin{examp} \label{#1}}
\newcommand{\eex}{\eop\end{examp}}
\newcommand{\bde}[1]{\begin{defi} \label{#1}}
\newcommand{\ede}{\end{defi}}
\newcommand{\bco}[1]{\begin{cor} \label{#1}}
\newcommand{\eco}{\end{cor}}
\newcommand{\bcorn}[2]{\begin{cor}[#1] \label{#1}}
\newcommand{\ecorn}{\end{cor}}
\newcommand{\bcon}[1]{\begin{con} \label{#1}}
\newcommand{\econ}{\end{con}}
\newcommand{\bfa}[1]{\begin{fact} \label{#1}}
\newcommand{\efa}{\end{fact}}
\newcommand{\bpr}[1]{\begin{problem} \label{#1}}
\newcommand{\epr}{\end{problem}}
\newcommand{\bprnn}[1]{\begin{problemnn} \label{#1}}
\newcommand{\eprnn}{\end{problemnn}}
\newcommand{\bprn}[2]{\begin{problem}[#1] \label{#2}}
\newcommand{\eprn}{\end{problem}}
\newcommand{\bexer}[1]{\begin{exercise} \label{#1}}
\newcommand{\eexer}{\end{exercise}}
\newcommand{\bre}[1]{\begin{rem} \label{#1}}
\newcommand{\ere}{\end{rem}}
\newcount\tpcnt
\newenvironment{tproblem}{%
  \global\advance\tpcnt1%
  \goodbreak\medskip\par\noindent\textbf{Problem~\the\tpcnt.}~}%
{%
  \goodbreak
}

\newenvironment{Solution}[1][]{%
  \goodbreak\smallskip\par\noindent\textbf{Solution{\if#1\empty\else~#1\fi}.}~}%
{%
  \goodbreak
}

\newenvironment{Proof}[1][]{
  \goodbreak\par
  \noindent\textit{Proof{\if#1\empty\else~#1\fi}.}~%
}{%
  \goodbreak\smallskip
}

\newcommand{\cR}{{\cal R}}

\newcommand{\bfc}{\mathbf{c}}

\newcommand{\bbF}{\mathbb{F}}

\newcommand{\Ff}{{\mathbb F}}

\newcounter{question_number}

\newenvironment{question}{\addtocounter{question_number}{1}\noindent{\bf Question \arabic{question_number}}}{\myaftermedspace}

\newenvironment{solution}{\noindent {\bf Solution:} \enspace}{\eop\myaftermedspace}

\newenvironment{hint}{\noindent {\bf Hint:} \enspace}{\eop\myaftermedspace}

\newenvironment{multisolution}[1]{\noindent {\bf Solution #1:} \enspace}{\eop\myaftermedspace}

\newcommand{\bqu}{\begin{question}}
\newcommand{\equ}{\end{question}}
\newcommand{\bs}{\begin{solution}}
\newcommand{\es}{\end{solution}}
\newcommand{\bh}{\begin{hint}}
\newcommand{\eh}{\end{hint}}
\newcommand{\bms}[1]{\begin{multisolution}{#1}}
\newcommand{\ems}{\end{multisolution}}

\newcommand{\btp}{\begin{tproblem}}
\newcommand{\etp}{\end{tproblem}}
\newcommand{\bts}{\begin{Solution}}
\newcommand{\ets}{\end{Solution}}

\newcommand{\bfr}{{\bf r}}
\newcommand{\bfs}{{\bf s}}

\newcounter{penumi}
\newenvironment{pit}{%
\begin{list}{(\roman{penumi})}{\usecounter{penumi}\setlength{\labelwidth}{1cm}\setlength{\itemindent}{0pt}\setlength{\topsep}{0pt}\setlength{\parsep}{0pt}\setlength{\partopsep}{0pt}\setlength{\itemsep}{0pt}}
} 
{\end{list}}
\newcommand{\bpit}{\begin{pit}}
\newcommand{\epit}{\end{pit}}

\newcommand{\bfaa}{\boldsymbol{a}}
\newcommand{\bfb}{{\bf b}}
\newcommand{\RRR}{{\bf R}}
\newcommand{\CCC}{{\bf C}}

\newcommand{\bfe}{{\bf e}}

\begin{document}
\title{Equal Requests are Asymptotically Hardest for Data Recovery}

\author{
  J\"uri Lember and Ago-Erik Riet\\
  Institute of Mathematics and Statistics\\ University of Tartu\\ 
                    Narva mnt 18, 51009 Tartu, Estonia\\
                    Email: \{juri.lember, ago-erik.riet\}@ut.ee
}

\maketitle

\begin{abstract}
   In a distributed storage system serving hot data, the data recovery performance becomes important, captured e.g.~by the service rate. We give partial evidence for it being hardest to serve a sequence of equal user requests (as in PIR coding regime) both for concrete and random user requests and server contents. 
   
   We prove that a constant request sequence is locally hardest to serve: If enough copies of each vector are stored in servers, then if a request sequence with all requests equal can be served then we can still serve it if a few requests are changed. 
   
   For random iid server contents, with number of data symbols constant (for simplicity) and the number of servers growing, we show that the maximum number of user requests we can serve divided by the number of servers we need approaches a limit almost surely. For uniform server contents, we show this limit is 1/2, both for sequences of copies of a fixed request and of any requests, so it is at least as hard to serve equal requests as any requests. For iid requests independent from the uniform server contents the limit is at least 1/2 and equal to 1/2 if requests are all equal to a fixed request almost surely, confirming the same.

   As a building block, we deduce from a 1952 result of Marshall Hall, Jr.~on abelian groups, that any collection of half as many requests as coded symbols in the doubled binary simplex code can be served by this code. This implies the fractional version of the Functional Batch Code Conjecture that allows half-servers.

\end{abstract}

\section{Introduction}

A linear code $C\le \Ff_q^n$ is determined by its linear encoder $G:\Ff_q^k\rightarrow \Ff_q^n$ which is a full-rank $k$-by-$n$ matrix over the finite field $\Ff_q$, known as the generator matrix, linearly mapping a row vector of $k$ data symbols of $\Ff_q$ to a row vector of $n$ coded symbols of $\Ff_q$ by multiplication. Data recovery codes, such as batch codes~\cite{Ishai} and PIR codes~\cite{fazelivardyyaakobi}, see~\cite{skachek} for a survey, in their classical as well as functional~\cite{zhangetzionyaakobi} or asynchronous~\cite{rietskachekthomas} versions, are exactly determined by a generator matrix up to a permutation of its columns. Thus these codes should in fact be called encoders. Batch codes were introduced for load-balancing in a distributed storage system~\cite{Ishai} and PIR codes, a generalization of batch codes (as PIR codes have weaker requirements), were introduced as a coding layer for multi-server information-theoretic private information retrieval (PIR) to reduce the storage overhead~\cite{fazelivardyyaakobi}. Here we only consider so-called ``primitive linear multiset batch codes'', where in particular ``linear'' implies that we use coding and not only replication.

%\section{Preliminaries}

For the usufulness of $G\in\text{Mat}_{k\times n}(\Ff_q)$, i.e.~a $k$-by-$n$ matrix $G$ over the finite field $\Ff_q$, as a code (encoder) optimized for recovery of data, different parameters have been studied. A general parameter is its service rate region~\cite{akr}\cite{akrs} that we will not discuss here in full generality, but which we can specify to the ``$t$-parameter'': the biggest $t$ such that any multiset/sequence of $t$ (linear combinations of) data symbols in $\Ff_q$ (user requests) can be recovered from pairwise disjoint sets of coded symbols in $\Ff_q$ (servers). What user requests we allow depends on the setup: whether the code is (functional) batch or PIR. 
For an asynchronous batch code user requests come in a stream and the system should serve a new one when any current (we do not know which) request has been served.

Serving a data symbol (user request) from a set a coded symbols (servers) under a linear model exactly corresponds to representing the standard basis vector (i.e.~unit vector) corresponding to the data symbol as a linear combination of the columns of $G$ corresponding to the coded symbols. We will thus also talk about serving a vector from a set of columns of $G$. Therefore the properties of our codes are exactly determined by the multiset of columns of $G$.

The matrix $G$ is defined to be a \emph{$t$-(functional) batch code}\cite{zhangetzionyaakobi} if we are able to serve any multiset of $t$ standard basis vectors (any vectors) from pairwise disjoint column sets of $G$. We will denote the largest $t$ such that $G$ is a $t$-(functional) batch code as $t_b(G)$ ($t_{fb}(G)$). For $G$ to be a \emph{$t$-(functional) PIR code}\cite{zhangetzionyaakobi} we must be able to serve $t$ copies of the same standard basis vector (the same vector), for any standard basis vector (any vector), and we denote such largest $t$ as $t_P(G)$ ($t_{fP}(G)$). 
The functional case corresponds to allowing linear combinations of data symbols as user requests. 
We say the matrix is $t$-PIR for $\bfr$ if we are able to serve $t$ copies of request $\bfr$ from it and write $t_{\bfr P}(G)$ for the largest such $t$. Thus $t_{fP}(G)=\min_{\bfr\ne{\bf0}}t_{\bfr P}(G)$.

We will call the $t_*(G)$ the \emph{$t$-parameter} of $G$ as a code where $*$ is one of $b,P,fb,fP,\bfr P$. We also write $t_*^{\le w}(G)$ if each column set in the service, i.e.~each \emph{recovery set} is required to have size $\le w$. The $t$-parameter does not depend on the permutation of the columns/coded symbols. Note the inequalities $t_{fb}(G)\le t_b(G)\le t_P(G)$ and $t_{fb}(G)\le t_{fP}(G)\le t_P(G)$, recall $t_{fP}(G)=\min_{\bfr\ne{\bf0}} t_{\bfr P}(G)$. All these remain true if we replace $t_*(G)$ by $t_*^{\le w}(G)$ for a fixed $w$.

In Sect.~\ref{buildingblocks} we develop the main building blocks. In Sect.~\ref{pirlocallyhardest} we prove if a constant request sequence can be served we can still serve it if a few requests are changed. Next we let $G\in\text{Mat}_{k\times n}(\Ff_2)$ be uniformly random throughout. In Sect.~\ref{randomgboundeddifferences} we prove concentration bounds for the $t$-parameters. In Sect.~\ref{asymptoticservicerateuniformg} we show we can asymptotically serve $\gamma n$ requests from $n$ columns where $\gamma=\frac12$, both if the requests are arbitrary or all equal. In Sect.~\ref{randomuserrequests} we show if we additionally let requests be random we can serve the fewest requests if they are equal to a fixed request almost surely, and note constant $\gamma$ exists in more general settings. In Sect.~\ref{conclusion} we conclude the paper.

\section{Building blocks}\label{buildingblocks}

If $G\in\text{Mat}_{k\times n}(\Ff_q)$ were uniformly random, from the law of large numbers, if $n$ grows much faster than $k$ we expect there to be around $n/q^k$ copies of each vector as a column of $G$. This motivates proving the following proposition that we will deduce from the result in~\cite{hall1952combinatorial}. For simplicity, from here on we assume $q=2$, but many results will in fact generalize. 

\begin{prop}\label{hallsimplex}
    Let $m$ be the largest integer such that $G$ contains as columns at least $2m$ copies of each vector of $\Ff_2^k$. Then $G$ is a $m2^k$-functional batch code, where moreover all recovery sets have size at most $2$, i.e.,~$t^{\le2}_{fb}(G)\ge m2^k$.
\end{prop}

This means that we can serve any sequence/multiset of $m2^k$ user requests (vectors from $\Ff_2^k\backslash\{{\bf0}\}$) if every $\bfe\in\Ff_2^k$ appears as a column of $G$ at least $2m$ times, where moreover each recovery set will have size at most $2$.

\begin{Proof}
    For general $m$, apply the $m=1$ case $m$ times, with column vectors as equally as possible and requests arbitrarily distributed into $m$ (almost) equal parts that are matched up.

   Now let $m=1$. Let the sequence of user requests be $R=(\bfr_1,\bfr_2\ldots,\bfr_{2^k})\in(\Ff_2^k)^{2^k}$. We will let $\bfs_1=-\sum_{i=2}^{2^k} \bfr_i$ in the abelian group $\Ff_2^k$ and $\bfs_i=\bfr_i$ for $i>1$. By the result in~\cite{hall1952combinatorial}, we can find two permutations $(\bfaa_i')_{i=1}^{2^k}$ and $(\bfb_i')_{i=1}^{2^k}$ of the abelian group $\Ff_2^k$, such that $\bfs_i=\bfaa_i'-\bfb_i'$ for each $i$. Let $\bfaa_i=\bfaa_i'+\bfc$ and $\bfb_i=\bfb_i'+\bfc$ for a fixed $\bfc$ such that now $\bfaa_1=\bfr_1$, also $\bfs_i=\bfaa_i-\bfb_i$ for each $i$, and also $(\bfaa_i)$ and $(\bfb_i)$ are permutations of $\Ff_2^k$. Finally, we can serve request $\bfr_1$ from column set $\{\bfaa_1\}$ and requests $\bfr_i$ for $i>1$ from column sets $\{\bfaa_i,\bfb_i\}$, using two copies of each vector of $
   \Ff_2^k\backslash\{\bfr_1-\bfs_1\}$ and one copy of $\bfr_1-\bfs_1$ in total from the columns of $G$.
\end{Proof}

\begin{rem}
   \begin{enumerate}
        \item The $m=1$ case of Proposition~\ref{hallsimplex} was proved in~\cite{yohananovyaakobi} Sect.~V by a technique similar to the more general~\cite{hall1952combinatorial} by Hall, Jr. from 1952. We point out~\cite{hall1952combinatorial} relatively easily implies the $m=1$ case of the Proposition.
        \item  In \cite{hall1952combinatorial} an explicit $\Theta((2^k)^2)$-time algorithm produces the permutations $(\bfaa_i),(\bfb_i)$ by  adding a new matching edge $\{\bfaa_j,\bfb_j\}$ and rewiring the previous ones at step $j\in[2^k]$. A similar algorithm in the special case was employed in~\cite{yohananovyaakobi}.
       \item The Proposition implies the fractional version of the Functional Batch Code Conjecture~\cite{kovacs2, yohananovyaakobi, hollmannkhathuriarietskachek, hollmannkhathuriarietskachek2, bgys} where we are allowed to split each column/server into `two halves'. The full Functional Batch Code Conjecture is open and seems to be hard to prove, and it is exactly the statement of the Proposition for $m=\frac12$. 
       \item     It does not necessarily provide the most efficient way to serve the requests. See Proposition~\ref{lowerbound} for an upper bound on the number of requests that can be served.
       \item    To serve request $\bf0$ we need no columns, and similarly, column $\bf0$ does not help to serve any request. Vector $\bf0$ was not treated differently here for ease of analysis.
   \end{enumerate}

\end{rem}

Let $N_G(\bfe)$ be the number of occurrences of $\bfe$ as a column of $G\in\text{Mat}_{k\times n}(\Ff_2)$ and let $N_R(\bfe)$ be the number of occurrences of $\bfe\in\Ff_2^k\backslash\{{\bf0}\}$ in user request sequence $R=(\bfr_1,\ldots,\bfr_v)\in(\Ff_2^k\backslash\{{\bf0}\})^v$. Let us denote $|G|=\sum_{\bfe\in\Ff_2^k}N_G(\bfe)=n$ and $|R|=\sum_{\bfe\in\Ff_2^k}N_R(\bfe)=v$ and let $|G\cap R|=\sum_{\bfe\in\Ff_2^k}\min\{N_G(\bfe),N_{R}(\bfe)\}$ for sizes as multisets.

\begin{prop}\label{lowerbound}
   If we can serve $R\in(\Ff_2^k\backslash\{{\bf0}\})^{v}$ from $G\in\text{Mat}_{k\times n}(\Ff_2)$, so $|G|=n$  and $|R|=v$, then 
\begin{equation}\label{canserve}
2|R|\le |G|-N_G({\bf0})+|G\cap R|.
\end{equation}  
\end{prop}
\begin{Proof}
  If $2|R|>|G|-N_G({\bf0})+|G\cap R|$ or equivalently $2(|R|-|G\cap R|)>|G|-N_G({\bf0})-|G\cap R|$ we can not serve $R$ from $G$ since for each request outside $G\cap R$ we need at least two nonzero columns of $G$ to serve it.  
\end{Proof}

\section{PIR is hardest among almost-PIR: local hardness}\label{pirlocallyhardest}

We have the intuition that many equal requests should be harder to serve than as many different requests. We can ask the following question.

\begin{ques}\label{conjecturehard}
     Is it true that $t_{fP}(G)=t_{fb}(G)$ for $G\in\text{Mat}_{k\times n}(\Ff_2)$? Perhaps even $t_{{\bfr}P}(G)=t_{fb}(G)$ if $N_G(\bfe)$  is minimized over $\bfe\in\Ff_2^k\backslash\{{\bf0}\}$ by $\bfr$?
\end{ques}

\begin{prop}
Answer to Question~\ref{conjecturehard} is `Yes' if $k=2$.
\end{prop}
\begin{Proof}
    We can without loss of generality, i.e., up to a symmetry which is multiplying each column by a fixed 2-by-2 invertible matrix over $\Ff_2$, assume $N_G((01))\le N_G((10)) \le N_G((11))$. Since for $k=2$ any inclusion-wise minimal recovery set has size at most 2, the following Lemma~\ref{disjoint} tells us that $t_{(01)P}(G)=t_{(10)P}(G)=N_G((01))+N_G((10))\le t_{(11)P}=N_G((01))+N_G((11))$, so $t_{fP}(G)=N_G((01))+N_G((10))$. To show $t_{fb}(G)=t_{fP}(G)=t_{(01)P}(G)$, i.e.~ that $N_G((01))+N_G((10))$ any requests can be served, we will greedily serve $|G\cap R|$ requests from recovery sets of size 1 and the rest from recovery sets of size 2. Enough recovery sets of size 2 remain for that: if $\Delta_{(01)}:=N_R((01))-N_G((01))>0$ then $\min\{N_G{((10))}-N_R{((10))}\,,\, N_G{((11))}-N_R{((11))}\}\ge\Delta_{(01)}$ and analogous relations hold for $\Delta_{(10)}$ and $\Delta_{(11)}$. 
\end{Proof} 

\begin{examp}\label{negative}
    Answer to Question~\ref{conjecturehard} is `No' already for $k=3$. Consider the matrix $G=\left(\begin{smallmatrix}1&1&1&1\\0&0&1&1\\0&1&0&1\end{smallmatrix}\right)
    $. Note that the sum of columns of $G$ is ${\bf0}$ and $\text{rank}(G)=3$, so $t_{fP}(G)=2$, as two copies of any $\bfr\in\Ff_2^3\backslash\{{\bf0}\}$ can be served from some subset of columns and its complement respectively. But $t_{fb}(G)\le t_b(G)=1$ as request sequence $((001),(010))$ cannot be served.
\end{examp}

The previous negative result shows the PIR coding regime is not at least as hard as the batch coding regime for some concrete matrices. In what follows we prove the positive results that it is hardest locally in the non-random case as well as almost surely in natural random models.

We now prove the lemmas towards our local hardness result.

\begin{lem}\label{disjoint}
    The distinct recovery sets (of nonzero columns) for a fixed $\bfr\in\Ff_2^k\backslash\{{\bf0}\}$  of size at most 2 are pairwise disjoint. In particular, the number $t_{\bfr P}^{\le2}(G)$ of pairwise disjoint such recovery sets for $G\in\text{Mat}_{k\times n}(\Ff_2)$ is their number which is \begin{equation}\label{rp} t_{\bfr P}^{\le2}(G)=N_G(\bfr)+\frac12\sum_{\bfe\in\Ff_2^k\backslash\{{\bf0},\bfr\}}\min\{N_G(\bfe),N_G(\bfr-\bfe)\}\end{equation} where the last sum counts each recovery set of size 2 twice.
\end{lem}

\begin{Proof}
    If $\bfr=\bfaa+\bfb=\bfaa'+\bfb'$ then $\{\bfaa,\bfb\}=\{\bfaa',\bfb'\}$ or $\{\bfaa,\bfb\}\cap\{\bfaa',\bfb'\}=\emptyset$ as $\bfaa-\bfaa'=\bfb'-\bfb$ and $\bfaa-\bfb'=\bfaa'-\bfb$. So pairs $\{\bfaa,\bfb\}$ with $\bfr=\bfaa+\bfb$ partition $\Ff_2^k$. The recovery set of size one is $\{\bfr\}$ which can be viewed as $\{\bfr,{\bf0}\}$ and vice versa. The assertion follows.
\end{Proof}

\begin{lem}\label{greedy}
    The maximum number of pairwise disjoint recovery sets for a fixed $\bfr\in\Ff_2^k\backslash\{{\bf0}\}$ is not reduced by choosing all recovery sets of size at most 2. That is, $$t_{\bfr P}(G)=t_{\bfr P}^{\le2}(G)+t_{\bfr P}(G')$$ where $N_{G'}(\bfr)=0$ and $N_{G'}(\bfe)=N_G(\bfe)-\min\{N_G(\bfe),N_G(\bfr-\bfe)\}$ for all $\bfe\in\Ff_2^k\backslash\{{\bf0},\bfr\}$, see also (\ref{rp}). Here $G'$ is what remains of $G$ after using the columns in all the recovery sets of size at most 2.
\end{lem}

\begin{Proof}
    Consider a collection of maximum possible size of pairwise disjoint recovery sets for $\bfr$, where the total size of all recovery sets used is minimum possible among such collections. Then if $A\subseteq B$ and $A$ and $B$ are both recovery sets for $\bfr$ we always use $A$ instead of $B$, i.e.~we only use minimal recovery sets. So we use all recovery sets of 1. There is no recovery set $A$ of size 2 which is not used where one of its columns is available and the other is used in a recovery set $B$ of size at least 3, as we could have used $A$ instead of $B$ and reduced the total size of used recovery sets. 
    
    Thus a recovery set of size 2 is not used only in the case where each of its columns is used in a different recovery set, which have to have size at least 3 by Lemma~\ref{disjoint}. 
Assume $A$ and $B$ are disjoint recovery sets of size at least 3 for $\bfr$, and $C\subseteq A\cup B$ is a recovery set for $\bfr$ with $|C|=2$ and $|A\cap C|=1=|B\cap C|$. Replace $A$ and $B$ by $C$ and $(A\cup B)\backslash C$ which is a recovery set for $\bfr$  as $\sum_{i\in (A\cup B)\backslash C}\bfc_i=\sum_{i\in A}\bfc_i+\sum_{i\in B}\bfc_i-\sum_{i\in C}\bfc_i=\bfr+\bfr-\bfr=\bfr$. 

By applying this step repeatedly the total number of used recovery sets and their total size do not change, and the number of used recovery sets of size at most 2 goes up. We will eventually have used all recovery sets of size at most $2$.
\end{Proof}

\begin{lem}\label{unique}
    Let $G\in\text{Mat}_{k\times n}(\Ff_2)$ with $N_G(\bfe)=2m$ for all $\bfe\in\Ff_2^k\backslash\{{\bf0}\}$. Then there is only one way (with equal columns indistinguishable) to serve $m2^k$ copies of a fixed $\bfr\in\Ff_2^k\backslash\{{\bf0}\}$, and this uses all the nonzero columns of $G$.
\end{lem}

\begin{Proof}
    Use $2m$ copies of each of the recovery sets $\{\bfr\}$ and $\{\bfe,\bfr-\bfe\}$ where $\bfe\in\Ff_2^k\backslash\{\bfr,{\bf0}\}$, i.e., $m2^k$ recovery sets in total. 
    Since each nonzero column of $G$ is used, and used in the unique possible recovery set for $\bfr$ of minimum size containing it, we have $t_{\bfr P}(G)=m2^k$, and this service is the unique service with $m2^k$ recovery sets.
\end{Proof}

Next we prove our main local hardness result partially confirming our intuition. 

\begin{teor}\label{locallyhard}
\emph{(``PIR is locally hardest'')}
    For an integer $m$ and vector $\bfr\in\Ff_2^k\backslash\{{\bf0}\}$ let $\mathcal{R}_\bfr^m\subseteq(\Ff_2^k\backslash\{{\bf0}\})^*$ be the family of the request sequences $R$ with at most $m2^k$ requests different from $\bfr$, and let $\mathcal{G}^m$ be the family of such matrices $G\in\text{Mat}_{k\times n}(\Ff_2)$ that contain $N_G(\bfe)\ge 2m$ copies of each $\bfe\in\Ff_2^k\backslash\{{\bf0}\}$ as a column. Then $G\in\mathcal{G}^m$ can serve
    any $R\in\mathcal{R}_\bfr^m$ with $|R|=t_{\bfr P}(G)$, the maximal length of a request sequence $(\bfr, \ldots, \bfr)$ that can be served by~$G$.
\end{teor}

This means that it is at least as hard to serve $v$ copies of request $\bfr$ by a matrix $G\in\mathcal{G}^m$ as it is to serve by $G$ any $R\in\mathcal{R}_\bfr^m$ consisting of $v$ requests.

\begin{Proof}
Let $H$ be the matrix obtained by removing $2m$ columns $\bfe$ from $G$ for all $\bfe\in \bbF_2^k\setminus \{{\bf 0}\}$, then $N_H(\bfe)=N_G(\bfe)-2m$ for all $\bfe\neq {\bf 0}$.
    
Now let $G'$ and $H'$ be the matrices obtained from $G$ and~$H$ by removing columns as in Lemma~\ref{greedy}. Then $N_{G'}(\bfr)=N_{H'}(\bfr)=0$, $N_{G'}(\bfe)=N_G(\bfe)-\min\{N_G(\bfe),N_G(\bfr-\bfe)\}$ for $\bfe\neq {\bf0}, \bfr$, and $N_{H'}(\bfe)=N_H(\bfe)-\min\{N_H(\bfe),N_H(\bfr-\bfe)\}=N_{G'}(\bfe)$ for $\bfe\neq {\bf0}, \bfr$. So we conclude that  $G'=H'$. 

Now using~(\ref{rp}), it is easily seen that $t_{\bfr P}^{\leq 2}(H)=t_{\bfr P}^{\leq 2}(G)-m2^k$ and by the above, we have that $t_{\bfr P}(G')=t_{\bfr P}(H')$. 

Then from Lemma~\ref{greedy}, we conclude that $t_{\bfr P}(H)=t_{\bfr P}(G)-m2^k$. 

Now let $R$ be a request sequence from $\cR^m_{\bfr}$ of length $t_{\bfr P}(G)$. Then $G$ can first serve $m2^k$ requests from $R$ including the requests different from $\bfr$, using Proposition~\ref{hallsimplex}, and then use the remaining columns $H$ to serve the remaining $t_{\bfr P}(G)-m2^k$ requests~$\bfr$, so $G$ can serve~$R$.
\end{Proof}

\section{Random $G$: bounded differences}\label{randomgboundeddifferences}

In what follows, we shall consider $G\in\text{Mat}_{k\times n}(\Ff_2)$ random with entries being i.i.d. Bernoulli with probability $\frac12$. Thus the columns of $G$ are the first $n$ vectors among i.i.d. random variables (abbreviated to r.v.'s) $\CCC_1,\CCC_2,\ldots$  with uniform distribution over $\Ff_2^k$. We shall refer to theses r.v.'s as the \emph{column process}. Then any $t$-parameter $t_*(G)$ is a function of  random columns, hence a r.v.. We shall denote these r.v.'s by $T_n^*=t_*(\CCC_1,\ldots, \CCC_n)$. It is rather easy to see that all these r.v.'s satisfy the \emph{bounded differences (BD)} property: changing one column of $G$ changes the value of $t_*(G)$   by at most 1. To see that take $*={\bfr P}$ and recall that $t_{\bfr P}(G)$ is the maximum number of copies of $\bfr$ the matrix $G$ can serve. 
Since each column belongs to at most one recovery set, by changing it, the maximum number of copies decreases or increases by at most one. By a similar argument the BD property can be checked for $t_{fP}(G)=\min_\bfr t_{\bfr P}(G)$ and $t_{fb}(G)$ and in fact for any of our $t_*(G)$ and $t_*^{\le w}(G)$. Since the columns are independent, by McDiarmid's inequality (see, e.g. \cite{lugosi}), for any $c>0$ we have $$\text{P}\left(T_n^*-\text{E}T_n^*\ge c\right)\le\exp\left(-\frac{2c^2}n\right).$$
McDiarmid's inequality allows in particular to obtain non-asymptotic confidence intervals, e.g.~with probability $\ge 1 - \alpha$ we have $T_n^*\ge\text{E}T_n^*-\sqrt{\frac{n}2\ln\frac1{\alpha}}.$ The use of this lower bound can be complicated in practice as $\text{E}T_n^*$ is generally not known. Then one can use more tractable suboptimal bounds such as $T_n^{*,\le 2}\le T_n^*$ which might have analytically known expectations, or estimate the expectation by simulations. For the upper bound, we use the \emph{superadditivity} property: for any $* = b,P,fb,fP,\bfr P$, any  $1\le n'<n$ and for any set of column vectors, we have 
 \begin{align*}
  t_*(\CCC_1,\ldots,\CCC_n)\ge t_*(\CCC_1,\ldots,\CCC_{n'})+t_*(\CCC_{n'+1},\ldots,\CCC_n).   
 \end{align*}
 The same holds for expectations, so  by Fekete's Lemma, 
 $ET^*_n/n\nearrow \sup_n ET^*_n/n$. In the next section, we shall show that the limit or supremum is 1/2, meaning that $ET^*_n\leq n/2$. Hence in the upper bound, $ET^*_n$ can be replaced by $n/2$ and the confidence interval reads: with probability $\ge 1 - \alpha$ we have $T_n^*\le\frac{n}2+\sqrt{\frac{n}2\ln\frac1{\alpha}}.$ Another use of McDiarmid's inequality is to obtain upper bounds for the absolute central moments: $$\text{E}\left|T_n^*-\text{E}T_n^*\right|^p\le p(n/2)^{p/2}\Gamma\left(\frac{p}2\right)$$
 for any $p>0$. For its derivation, see, e.g.~\cite{LemberHoudre}. Setting $p=2$, we obtain $\text{Var}(T_n^*)\le n$.

\begin{rem}
    McDiarmid's inequality and its consequences above only use the BD property and that columns of $G$ are independent. Hence the entries within one column can be dependent and the columns need not be identically distributed.
\end{rem}

%----------------------------------------
\section{Asymptotic service rate for uniform $G$ is lowest for constant request sequences}\label{asymptoticservicerateuniformg}

Recall that the r.v.'s $T^*_n$ satisfy the following inequalities: $T_n^{fb,\le2}\leq T_n^{fb}\leq T_n^{f  P}\leq  T_n^{ \bfr P}$. We shall now show that when divided by $n$, all these sequences converge almost surely (abbreviated to a.s., i.e.~with probability 1) to 1/2. This means that asymptotically serving any constant request sequence is as hard as serving all sequences. It also implies that the total number of all recovery sets with sizes more than two that we need is asymptotically negligible. In other words, for large $n$ we can serve any request with recovery sets consisting of one or two elements, without losing much.

\begin{teor}\label{asymptotic} For any fixed $\bfr\in\Ff_2^k\backslash\{{\bf0}\}$ we have
$$\lim_n\frac{T_n^{fb,\le2}}{n}=\lim_n\frac{T_n^{\bfr P}}{n}=\frac12$$ a.s..
\end{teor}

%\end{proof}

\begin{Proof}
Recall that $N_G(\bfe)$ is the number of occurrences of $\bfe$ as a column of $G$. From Hoeffdings's inequality (see, e.g, \cite{lugosi}) we have $$\text{P}\left(\frac{N_G(\bfe)}{n}-\frac1{2^k}\ge\varepsilon\right)\le\exp(-2n\varepsilon^2).$$

Let $\varepsilon_n\searrow0$ slowly enough s.t.~$\sum_n\exp(-2n\varepsilon_n^2)<\infty$. By the Borel-Cantelli Lemma \begin{equation}\label{borelcantelli}    \text{P}\left(\frac{N_G(\bfe)}{n}\ge\frac1{2^k}-\varepsilon_n \text{\;for all\;}\bfe\in\Ff_2^k\text{\;eventually}\right)=1.
\end{equation}
We will abbreviate eventually to ev..

Denote $m_n:=n\left(1/{2^{k+1}}-{\varepsilon_n}/2\right)$. 

Thus if ${N_G(\bfe)}/n\ge1/{2^k}-\varepsilon_n$ for all $\bfe\in\Ff_2^k$ then we have $N_G(\bfe)\ge2m_n$ for all $\bfe\in\Ff_2^k$ and in that case we deduce from Proposition~\ref{hallsimplex} that $$T_n^{fb,\leq 2}\ge m_n2^k=n\left(\frac12-2^{k-1}\varepsilon_n\right),$$ 
thus $$\frac{T_n^{fb,\leq 2}}n\ge\frac12-2^{k-1}\varepsilon_n.$$

From~(\ref{borelcantelli}) we have $\text{P}\left({T_n^{fb,\leq 2}}/n\ge\frac12-2^{k-1}\varepsilon_n\text{\;\;ev.}\right)=1$, thus $$\liminf_n\frac{T_n^{fb,\leq 2}}n\ge\frac12, \quad {\rm a.s.}$$
%-----------------------------------------------------------
To prove the upper bound, fix $\bfr$ and observe  that from   (\ref{canserve}), it follows that
$$2T_n^{\bfr P}\leq n-N_G(0)+\min\{N_G(\bfr), T_n^{\bfr P}\}\leq n+N_G(\bfr)-N_G(\bf0),$$
because $|G\cap R|=\min\{N_G(\bfr),T_n^{\bfr P}\}$ when the request sequence $R$ consists of $T_n^{\bfr P}$ copies of $\bfr$. 
Hence
$${T_n^{rP}\over n}\leq {1\over 2}+{N_G(\bfr)-N_G(0)\over 2n}\to {1\over 2},\quad {\rm a.s.},$$
Thus $$\limsup_n {T_n^{rP}\over n}\leq 1/2.$$

\end{Proof}

\begin{rem}
    The lower bound followed from Proposition~\ref{hallsimplex} and the concentration of the frequencies of vectors as columns. For a slightly weaker upper bound $\limsup_n({T_n^P}/n)\le\frac12$ we could also use that $t_P(G)\le\text{d}(C_G)$, where $\text{d}(C_G)$ is the minimum distance of the code generated by $G$, i.e.~row span of $G$ (see~\cite{lipmaaskachek, hollmannluhaaar}), and that $\limsup_n(\text{d}(C_G)/n)\le\frac12$, a.s.
\end{rem}

\section{Random user requests}\label{randomuserrequests}

We now generalize our model by letting requests be random as well. To be more precise, let
 the requests be i.i.d. random vectors $\RRR_1,\RRR_2,\ldots$  taking values in  $\Ff_2^k\backslash\{{\bf0}\}$, we will refer to them as the \emph{request process}. The distribution $Q$ of $\RRR_i$ will be referred to as the \emph{request distribution}. Given $G\in\text{Mat}_{k\times n}(\Ff_2)$, let the r.v. $$V(G)=\max\{V:\{\RRR_1,\ldots,\RRR_V\}\text{\;can be served by\;} G\}.$$ Thus $\RRR_1,\ldots,\RRR_V$ can be served but $\RRR_1,\ldots,\RRR_{V+1}$ can not be. As in the previous sections, we consider $G$ also random with i.i.d. uniformly distributed columns and independent from the request process. Let $V_n=V(G)$. Hence $V_n$ is a function depending on $n$ random columns and at most $n$ random requests. Observe that when $\RRR_i=\bfr$ a.s., i.e.~the request distribution $Q$ is degenerate ($Q=\delta_{\bfr}$), then $V_n=T_n^{\bfr P}$. Clearly for any $Q$ we have  $V_n\geq T_n^{fb}$, so by Theorem~\ref{asymptotic} we have $\liminf_n(V_n/n)\ge\frac12$.

We now argue that when $Q$ is not degenerate, then $\lim_n({V_n}/n)$, if it exists, might be bigger than $\frac12$. This confirms the  fact that asymptotically, equal requests are hardest to serve, i.e.~allowing randomness in the request process increases the asymptotic proportion of served requests. We illustrate it for the case of the uniform $Q$, i.e. P$(\RRR_i=\bfe)=1/({2^k-1})$ for all $\bfe\in\Ff_2^k\backslash\{{\bf0}\}$. Recall that  $\bf0$ is not allowed as a request, so such a $Q$ has maximum entropy. 

\begin{prop} For  $Q$ uniform,  $V_n/n\to 1-2^{-k}$, a.s.. \end{prop}

\begin{Proof}
Let $0<\alpha<1-2^{-k}$, hence ${\alpha}/({2^k-1})<1/{2^k}$. For every $\bfe\ne{\bf0}$ let $N_R^\alpha(\bfe)=\sum_{i=1}^{n\alpha}I_\bfe(\RRR_i)$ be the number of occurrences of $\bfe$ among the first $n\alpha$ requests. Clearly $N_R^\alpha(\bfe)/(n\alpha)\rightarrow 1/(2^k-1)$ a.s. Take $\varepsilon>0$ small enough such that $\varepsilon<1/2^k-\alpha/(2^k-1)$, thus 
\begin{equation}\label{eps}
\alpha/(2^k-1)+\alpha\varepsilon/2<1/2^k-\varepsilon/2.    
\end{equation}
Clearly, as $G$ is uniform and $n\rightarrow\infty$, (``ev.'' is ``eventually'')
\begin{align*}
&\text{P}\Big(\max_{\bfe\ne{\bf0}}N_R^\alpha(\bfe)<\big(\frac\alpha{2^k-1}+\frac{\alpha\varepsilon}2\big) n,\text{\;ev.}\Big)=1,  \\
&\text{P}\Big(\min_{\bfe\ne{\bf0}}N_G(\bfe)>\big(\frac1{2^k}-\frac\varepsilon2\big) n,\text{\;ev.}\Big)=1.
\end{align*}
 Hence by (\ref{eps}) $$\text{P}\left(\sum_{\bfe\ne{\bf0}}\min\{N_G(\bfe),N_R(\bfe)\}=\sum_{\bfe\ne{\bf0}}N_R^\alpha(\bfe)=\alpha n,\text{\;ev.}\right)=1.$$ So $\text{P}(V_n\ge\alpha n,\text{\;ev.})=1$, i.e. $\liminf_n(V_n/n)\ge\alpha$, a.s. This holds for any $\alpha<1-2^{-k}$ implying that $\liminf_n(V_n/n)\ge 1-2^{-k}$ a.s.

For the upper bound, argue similarly: take $\alpha>1-2^{-k}=(2^k-1)/2^k$, hence $\alpha/(2^k-1)>1/2^k$. For any $\varepsilon>0$ and any $\bfe\ne{\bf0}$ we have $N_R^\alpha(\bfe)>\left(\alpha/(2^k-1)-\varepsilon\right)n$, eventually. On the other hand $N_G(\bfe)<(1/2^k+\varepsilon)n$, eventually. Hence for every $n$ big enough not all $\alpha n$ requests can be served, so that $V_n<\alpha n$, eventually. Thus $\limsup_n(V_n/n)\le\alpha$, a.s. This holds for any $\alpha>1-2^{-k}$, so $\limsup_n(V_n/n)\le 1-2^{-k}$, a.s. This completes the proof.
\end{Proof}

For uniform $G$, we could also ask about the maximum number of requests among the given requests that can be served by it. Given an i.i.d. request process $\RRR_1,\RRR_2,\ldots,$ with $\RRR_i\sim Q$ and a matrix $G$, let $L(G;\RRR_1,\ldots,\RRR_n)$ be the maximum size of a subset $I\subseteq[n]$ such that the request sequence $\{\RRR_i\,|\,i\in I\}$ can be served by $G$. Clearly $t_{fb}(G)\leq V(G)\le L(G)\le n$. Recall that the random  $G$ is just the first $n$ elements of the i.i.d.~uniform column process $\CCC_1,\CCC_2,\ldots$, so $L_n:=L(G)=L(\CCC_1,\ldots,\CCC_n;\RRR_1,\ldots,\RRR_n)$. Recall the column and request processes are independent. We obviously have superadditivity: for any $1\le n'<n$ we have 
\begin{align*}
& L(\CCC_1,\ldots,\CCC_n;\RRR_1,\ldots,\RRR_n)\ge L(\CCC_1,\ldots,\CCC_{n'};\RRR_1,\ldots,\RRR_{n'})\\
 &+L(\CCC_{n'+1},\ldots,\CCC_n;\RRR_{n'+1},\ldots,\RRR_n),   
\end{align*}
 so by Kingman's Subadditive Ergodic Theorem (see, e.g. \cite{durrett}) there is a constant $\gamma\in\left[\frac12,1\right]$ such that $L_n/n\rightarrow\gamma$, a.s. and $\text{E}(L_n)/n\rightarrow\gamma$. It is not hard to see that 
if $Q$ is uniform, then $\gamma=1-2^{-k}$, just like above. Indeed, $\gamma\geq 1-2^{-k}$ follows from the fact that $L_n\geq V_n$. To obtain the upper bound, take $\alpha>1-2^{-k}$ and note that a necessary condition for $L_n\geq \alpha n$ is that $n-N_G({\bf0})=\sum_{\bfe\ne \bf0}N_G(\bfe)\geq \alpha n$. Thus $N_G({\bf0})/n>(1-\alpha)$ implies $L_n<\alpha n$. Since $1-\alpha<2^{-k},$ it follows that $\text{P}(L_n/n< \alpha)\geq \text{P}\big(N_G({\bf0})/n>(1-\alpha)\big)\to 1.$ Therefore $\gamma\leq \alpha.$
 
 It is also easy to see that $L_n$ satisfies the BD property, unlike $V_n$. Since $L_n$ is a function of $2n$ independent random variables, (two-sided) McDiarmid's inequality in this case reads $$\text{P}\left(|L_n-\text{E}L_n|>c\right)\le 2\exp\left(-\frac{c^2}n\right).$$
The limit $\gamma$ in Kingman's theorem depends on the request distribution $Q$ and on the column distribution $P$. In our paper $P$ is uniform over $\Ff_2^k$, but it need not necessarily be so.  We have shown that when $P$ is uniform, then $\gamma(P,Q)\geq 1/2$ and when, in addition, $Q=\delta_{\bfr}$ (degenerate), then $\gamma(P,Q)=1/2$ (PIR is the hardest). We have also shown that when $P$ and $Q$ are both uniform ($Q$ over non-zero vectors), then $\gamma(P,Q)=1-2^{-k}.$ The other cases are open and determining the function $\gamma(P,Q)$ is a challenge for future research. Kingman's theorem in fact holds for any ergodic process $(\CCC_1,\RRR_1),(\CCC_2,\RRR_2),\ldots,$ i.e.~the column and request process need not necessarily be i.i.d.~or mutually independent. So, for example the limit $\gamma$ also exists when the request process and the column process are both (possibly dependent)  stationary Markov chains.

\section{Conclusion}\label{conclusion}

We consider the question if equal user requests are hardest to serve for a distributed storage/service system, like a batch code or a PIR code. For concrete, non-random $k$-by-$n$ generator matrices $G$ over $\Ff_2$ we prove that serving a request sequence $\bfr,\ldots,\bfr$ of $v$ requests is at least as hard as serving any request sequence in a Hamming ball around this sequence, that is, it is locally hardest. The radius of the Hamming ball depends on the matrix $G$, concretely on the minimum number of occurrences of $\bfe$ as a column of $G$, over all $\bfe\ne{\bf0}$.

But there are concrete matrices $G$ already with $3$ rows which can serve two copies of any request, but for which there is a pair of two different requests that cannot be served by $G$. 

For uniformly random generator matrices we prove serving $\bfr,\ldots,\bfr$ for any fixed $\bfr\ne{\bf0}$ is indeed hardest asymptotically, with $k$ fixed and  $n\rightarrow\infty$, and we determine that the asymptotic service rate, i.e.~the ratio of the number of requests that can be served to the number of servers needed, is $\gamma=1/2$. Letting moreover requests be i.i.d. and independent from $G$, the hardest case is again where the requests are all equal to a fixed request with probability 1. On the other hand with requests also uniform but over nonzero vectors, $\gamma=1-2^{-k}$.

An asymptotic service rate $\gamma=\gamma(P,Q)$ exists for any i.i.d. requests (with distribution $Q$) and any i.i.d. server contents (with distribution $P$) independent from the requests, and in more general settings. Its determination in other cases is open.

We explain how to estimate the respective random variables, give confidence intervals, and estimate their moments.

As a building block, we deduce from a 1952 theorem of Marshall Hall, Jr.~on abelian groups that the doubled binary simplex code can serve any request sequence of half its length plus one. This easily implies the fractional version of the Functional Batch Code Conjecture where half-servers are allowed.

\section*{Acknowledgment}

We would like to thank Kristiina Oksner and Vitaly Skachek for Example~\ref{negative} and Henk D.L. Hollmann for a better derivation of Theorem~\ref{locallyhard} from the preceding Lemmas. We would like to thank them all, Tomas Ju\v{s}kevi\v{c}ius and Junming Ke for useful discussions. This work was supported by the Estonian Research Council through research grant PRG865.

\end{document}